\documentclass[amsmath,floatfix,twocolumn,superscriptaddress]{revtex4-1}
\usepackage{amssymb,amsmath} \usepackage{graphicx,array}
\usepackage{amsfonts} \usepackage{dcolumn} \usepackage{psfrag}
\usepackage{bm} \usepackage{color}

\begin{document}
\title{Polymorphism of Bulk Boron Nitride}

\author{Claudio Cazorla} \affiliation{School of Materials Science and
  Engineering, UNSW Australia, Sydney NSW 2052, Australia}

\author{Tim Gould} \affiliation{Queensland Micro- and Nanotechnology
  Centre, Griffith University, Nathan, QLD 4111, Australia}
\affiliation{School of Environment and Science, Griffith University,
  Nathan, QLD 4111, Australia}

\maketitle

{\bf Boron nitride (BN) is a material with outstanding technological
  promise because of its exceptional thermochemical stability,
  structural, electronic and thermal conductivity properties, and
  extreme hardness. Yet, the relative thermodynamic stability of its
  most common polymorphs (diamond-like cubic and graphite-like
  hexagonal) has not been resolved satisfactorily because of the
  crucial role played by kinetic factors in the formation of BN phases
  at high temperatures and pressures (experiments), and by competing
  bonding, electrostatic and many-body dispersion forces
  in BN cohesion (theory). This lack of understanding hampers the
  development of potential technological applications, and challenges
  the boundaries of fundamental science.  Here, we use high-level
  first-principles theories that correctly reproduce all important
  electronic interactions (the adiabatic-connection
  fluctuation-dissipation theorem in the random phase approximation)
  to estimate with unprecedented accuracy the energy differences
  between BN polymorphs, and thus overcome the accuracy hurdle
  that hindered previous theoretical studies. We show that
  the ground-state phase of BN is cubic and that the frequently
  observed hexagonal polymorph becomes entropically
  stabilized over the cubic at temperatures slightly above ambient
  conditions ($T_{\rm c \to h} = 335 \pm 30$~K).
  We also reveal a low-symmetry monoclinic phase
  that is extremely competitive with the other low-energy polymorphs
  and which could explain the origins of the experimentally observed
  ``compressed h--BN'' phase. Our theoretical findings therefore
  should stimulate new experimental efforts in bulk BN as well as
  promote the use of high-level theories in modelling of
  technologically relevant van der Waals materials.}\\

\begin{figure}[h]
  \centerline{
    \includegraphics[width=89mm]{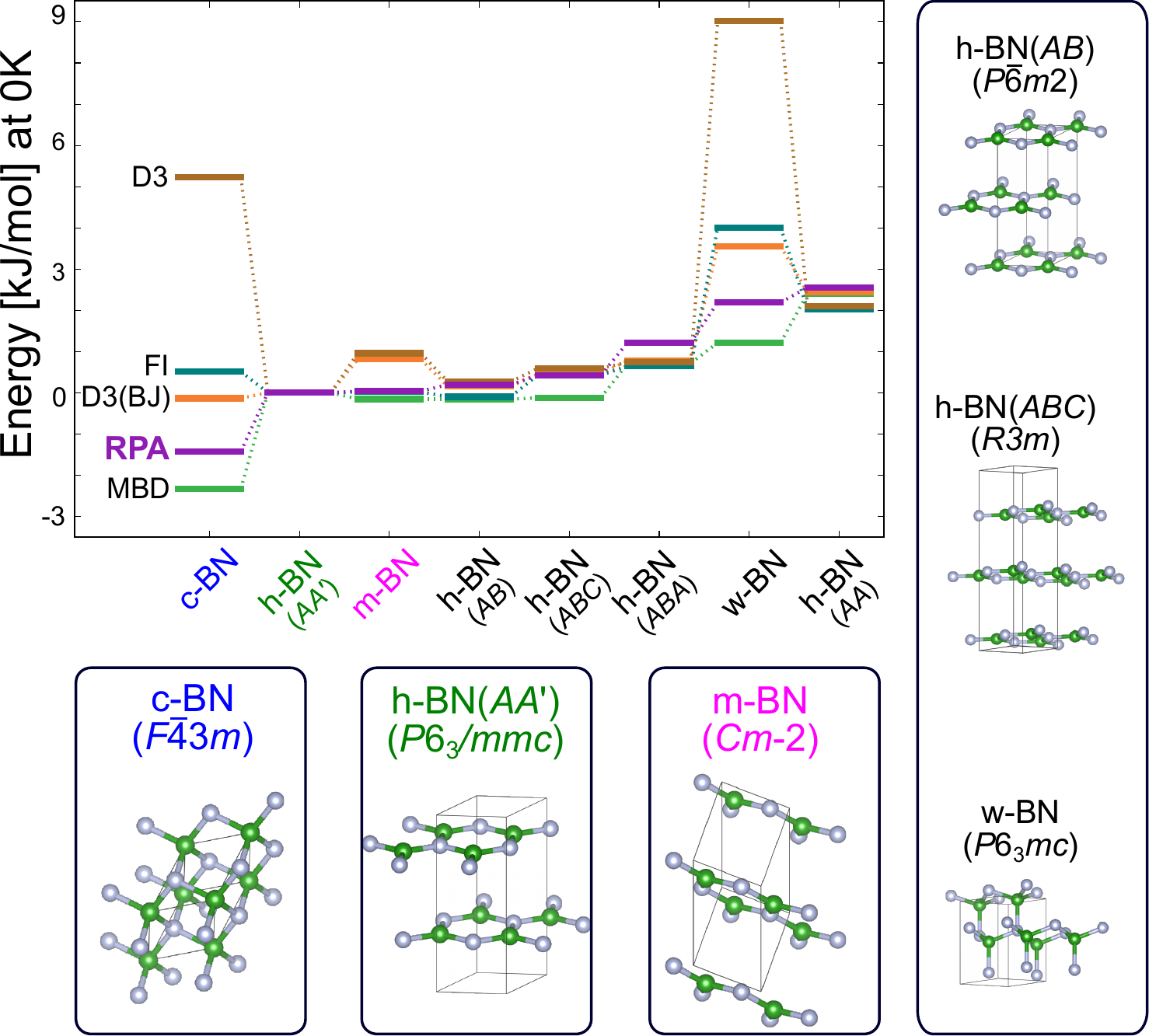}}
  \caption{{\bf BN polymorphs and corresponding zero-temperature
      energies calculated with first-principles based methods.}~The energy of
    the h--BN$(AA')$ polymorph is taken as the reference value in each
    of the series. ``D3'' \cite{d31,d32}, ``FI'' \cite{fi}, ``D3(BJ)''
    \cite{d3bj}, ``RPA'' \cite{rpa0,rpa1,rpa2,rpa3,rpa4}, and
    ``MBD'' \cite{mbd} stand for different dispersion corrected
    first-principles methods, all taken over PBE \cite{pbe}
      as a base functional. LDA \cite{lda} and PBE
    methods provide results outside the selected range (Supplementary
    Table~2). Energy results include quantum nuclear effects through
    zero-point energy corrections as calculated with the LDA method
    (Methods). The notation used to refer to the BN polymorphs
    throughout the text along with the corresponding space groups and
    crystal structures are specified; letters within parentheses
    accompanying the hexagonal polymorphs indicate the stacking sequence
    between consecutive B--N planes along the hexagonal $c$--axis.}
  \label{fig0K}
\end{figure}

In spite of the tremendous technological interest of bulk boron
nitride (BN) \cite{dean10,tang02}, fundamental knowledge of its 
phase diagram remains contentious to this day. The two most common BN polymorphs 
possess hexagonal (h--BN) and cubic (c--BN) symmetries and are structurally
analogous to the graphite and diamond phases of carbon, respectively
(Fig.~\ref{fig0K}). Based on empirical observations, and in analogy to
the carbon phase diagram, h--BN generally is regarded as the most
stable BN polymorph at ambient conditions
\cite{pease50,corrigan75,narayan16}: c--BN does not exist in nature
and its synthesis in laboratories requires high-temperature and
high-pressure conditions. Strikingly, experimental phase
diagrams based on thermodynamic and {\em in situ} x-ray diffraction
measurements strongly suggest that c--BN is more stable than h--BN at
normal conditions \cite{bundy62,solozhenko99,will00}.  The reported
c--BN$\leftrightarrow$h--BN transition temperatures as extrapolated to
ambient pressure, however, vary by as much as
$420 < T^{\rm exp}_{\rm c \to h} < 1500$~K \cite{will00}. The
cause of this huge variation is the critical importance of kinetic
effects on the c--BN$\leftrightarrow$h--BN transformation, which
depends strongly on difficult-to-control parameters like grain size,
defects concentration, and the purity of the starting
material \cite{solozhenko99,will00}.

Calculations based on quantum mechanics
are free of the abovementioned kinetic
factors affecting experiments; however, weakly bound layered
materials, like most BN polymorphs, are known to pose serious
challenges to standard first-principles methods which do not include
dispersion (van der Waals) interactions (e.g., density
functional theory -- DFT -- based on the local density -- LDA -- 
\cite{lda} and generalized gradient -- GGA -- \cite{pbe}
approximations to the exchange-correlation energy). In
fact, DFT estimations based on LDA and GGA reach
contrary conclusions on the relative stability of the c--BN and h--BN
polymorphs (see, for instance, \cite{kresse99,ahmed07}), thus adding
further confusion to the BN phase diagram puzzle. The last decade has
seen extraordinary progress in the development of dispersion-corrected
theories that overcome the limitations of standard theories; modern
approaches like the D3 empirical correction \cite{d31,d32,d3bj} and
many-body dispersion methods \cite{mbd,fi} are able to reliably
optimize complex layered polymorph structures, and predict energy
differences between them with fair accuracy \cite{tawfik2018}.
But even these recent developments might not be relied upon
to determine the relative thermodynamic stability of low-energy BN
polymorphs since the involved energy differences can be below $\sim
1$~kJ/mol ($\sim 10$~meV per formula unit), that is, the
characteristic scale of non-systematic errors in most dispersion
approximations.

  The challenge of capturing these small energy differences
  is made more difficult by the importance of many-body
  interactions in dispersion bound systems. Low-dimensional systems,
  like the layered polymorphs of BN, exhibit collective dispersion
  interactions \cite{mbd,ambrosetti2016,hermann2017} identified as
  ``Type~B'' non-additivity by Dobson \cite{dobson2014}.
  Type-B effects cannot be represented as a sum over pairwise
  interactions, as is done in many dispersion correction schemes, or
  even in more sophisticated perturbation approaches like
  M{\o}ller-Plesset theory.
  This makes BN polymorph ranking especially challenging, as any
  theoretical model must incorporate Type-B terms in its underlying
  physics.

Here, we use the adiabatic-connection fluctuation-dissipation theorem
in the random-phase approximation (RPA), a method that has been shown
to reliably and seamlessly treat both strong and weak interactions in
a wide variety of systems, and which has a full treatment of
many-body interactions \cite{rpa0,rpa1,rpa2,rpa3,rpa4}. We thereby
calculate the energies of BN polymorphs with unprecedented accuracy
to overcome previous theoretical bottlenecks. We show that the
ground-state of BN is c--BN and that the h--BN polymorph becomes
thermodynamically most stable at temperatures close to ambient,
namely, $T_{\rm c \to h} = 335 \pm 30$~K.  By using a multi-stage
approach (Methods, Supplementary Methods, and Supplementary Table~1) to 
determine and rank low energy structures from an initial list of 15 structures, 
we reveal a low-symmetry monoclinic phase that turns out to be
energetically very competitive with other known polymorphs, and which
could explain the origins of previously overlooked experimental
observations. We discuss the causes of the phase stability phenomena
revealed in bulk BN, and argue the need for using high-level theories
in computational studies of technologically relevant materials.

Figure~\ref{fig0K} shows the zero-temperature energies of most popular
BN polymorphs relative to that of h--BN$(AA')$, calculated with
several dispersion-corrected DFT methods and RPA (corresponding space
groups and polymorph notation employed throughout the text are
explained therein). These zero-temperature energies include quantum
nuclear effects \cite{cazorla17} via zero-point energy (ZPE)
corrections (Methods), although the ZPE effects do not lead to
qualitative changes in the relative stability of BN polymorphs
(Supplementary Table~2). Our results demonstrate two key
points. Firstly, the RPA ordering of low-energy states confirms that
c--BN is the ground state \cite{bundy62,solozhenko99,will00}, which is
about $1$--$2$~kJ/mol 
lower in energy than the h--BN$(AA')$, m--BN (discussed below),
h--BN$(AB)$, and h--BN$(ABC)$ (also known as r--BN) polymorphs. And
secondly, the many-body dispersion (MBD) method \cite{mbd} agrees quite well with RPA, 
this being the only semi-empirical method that predicts correctly the
energy ordering among all the most competitive phases. Although,
the D3 method with Becke-Johnson damping [D3(BJ)] \cite{d3bj}, and
the fractionally ionic MBD (FI) \cite{fi} method come close.
Energy results obtained with the LDA and PBE methods turn
out to be of unsatisfactory quality (Supplementary Table~2).

\begin{figure}[h]
  \centerline{
    \includegraphics[width=89mm]{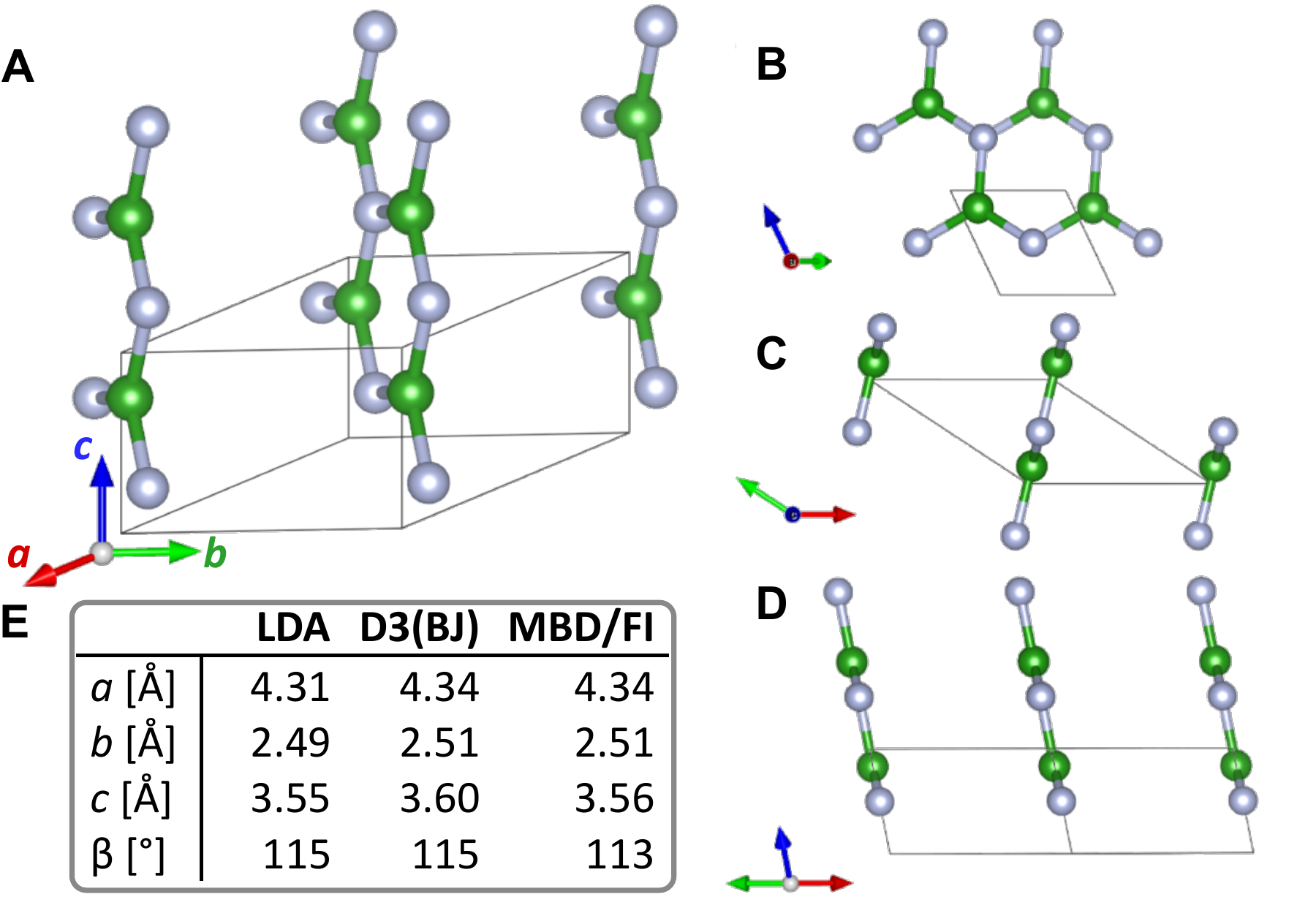}}
  \caption{{\bf The monoclinic phase m--BN (space group $Cm$) reported in this study.} 
    {\bf A-D}~Projections showing its similarities to the h--BN polymorph;
    {\bf E}~Key structural parameters found via different methods (Supplementary Table~3). 
    Boron and nitrogen atoms are represented with green and blue spheres, respectively.}
  \label{figM}
\end{figure}
Our study reveals a low--symmetry monoclinic phase, denoted here
as m--BN (space group $Cm$, different from a previously predicted
monoclinic phase \cite{zhang13} hence the ``2'' in Fig.~\ref{fig0K}),
that is energetically very competitive with respect to the h--BN
polymorphs ($\Delta E^{\rm RPA} < 1$~kJ/mol). This new phase is
vibrationally stable and presents a reduced two--atoms unit cell with
equilibrium parameters $a_{m} = 4.34$, $b_{m} = 2.51$, $c_{m} =
3.56$~\AA~, and $\beta = 113^{\circ}$, as obtained with MBD and FI
methods (Figure~\ref{figM}, Supplementary Table~3, and Supplementary
Figure~1). The predicted m--BN polymorph is structurally very similar
to a previously reported monoclinic phase that was experimentally
observed during the c--BN$\leftrightarrow$h--BN transformation
occurring at high--$P$ high--$T$ conditions
($a^{\rm exp}_{m} = 4.33$, $b^{\rm exp}_{m} =
2.50$, $c^{\rm exp}_{m} = 3.1$--$3.3$~\AA~, and $\beta^{\rm exp} =
92$--$95^{\circ}$) and which was named as ``compressed h--BN''
\cite{horiuchi95}; hence, we tentatively identify the two phases as
the same (up to the non-negligible effects of pressure and temperature
disregarded in our simulations).

\begin{table}[h]
  \begin{ruledtabular}\begin{tabular}{llrrrrrrr}
      && $C_{11}$ & $C_{22}$ & $C_{33}$ &
      $C_{12}$ & $C_{23}$ & $C_{13}$ & $B_{VRH}$ \\\hline
      c--BN & LDA &
       997 &  997 &  997 & 101 & 101 & 101 & 402 \\
      c--BN & RPA &
       968 &  968 &  968 &  81 &  81 &  81 & 378 \\\hline
      h--BN($AA'$) & LDA &
       923 &  926 &   28 & 174 &   3 &   3 & 138 \\
      h--BN($AA'$) & RPA &
       910 &  915 &   29 & 153 &  -4 &  -5 & 131 \\\hline
      m--BN & LDA &
       751 &  839 &   45 & 156 & 100 &  68 & 160 \\
      m--BN & RPA &
       745 &  830 &   43 & 137 &  90 &  58 & 151 \\
  \end{tabular}\end{ruledtabular}
  \caption{Elastic constants and bulk modulus in the Voigt-Reuss-Hill
    approximation (``VRH'', as this is appropriate for polycrystalline
    samples \cite{ivanovskii12}) of the most stable
    polymorphs calculated with the LDA \cite{lda} and
    RPA \cite{rpa0,rpa1,rpa2,rpa3,rpa4} methods. LDA results are
    very close to those from RPA, showing the suitability of LDA for 
    estimating second energy derivatives. Results are in units of GPa.}
  \label{tab1}
\end{table}
  
In Table~\ref{tab1}, we show the elastic constants, $C_{ij}$
(given in Voigt notation), of the 
low energy polymorphs calculated with the RPA and LDA methods. The
elastic properties of the monoclinic phase are very similar to those
of h--BN$(AA')$, a result that along with the minute energy
difference among the two might explain the causes of the
experimentally observed ``compressed h--BN''/h--BN coexistence
\cite{horiuchi95} (Supplementary Table~4). 
We note that the series of $C_{ij}$ values
estimated with the LDA and RPA methods are in very good agreement,
with key LDA values systematically 5--6\% above their RPA
counterparts. This outcome supports the use of LDA for assessing the
vibrational properties of BN polymorphs accurately, a task which has
a prohibitive computational cost in RPA.

\begin{figure}[h]
  \centerline{\includegraphics[width=89mm]{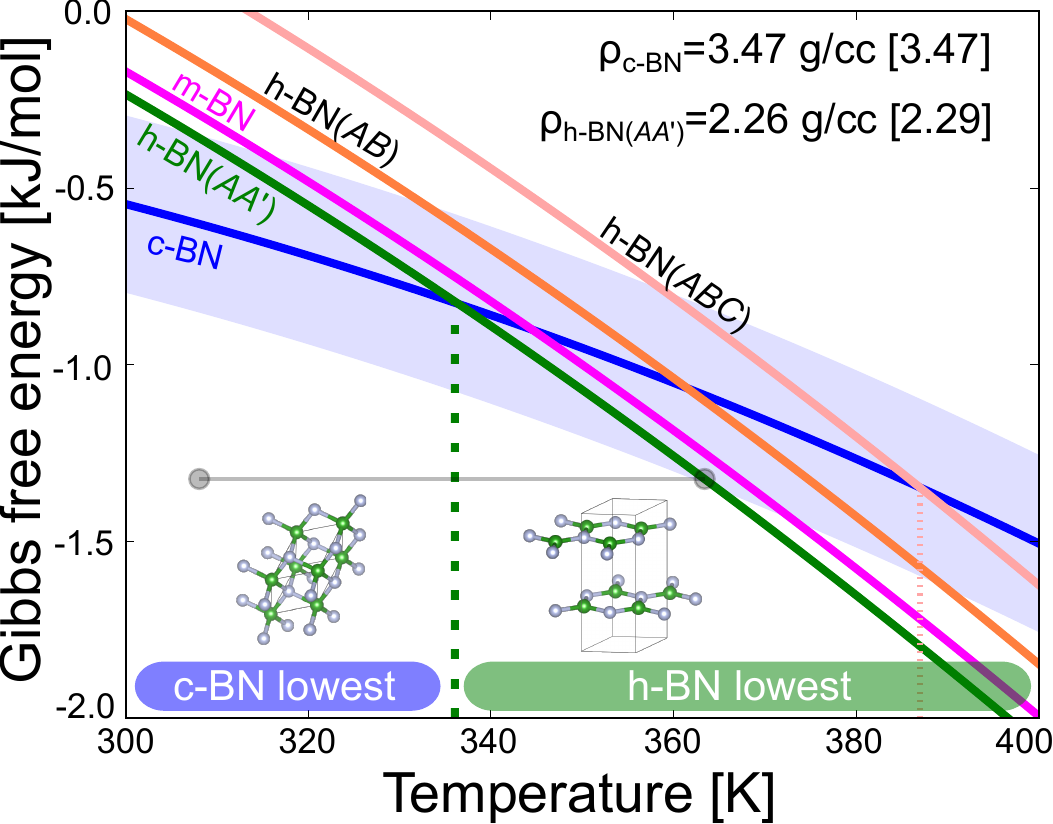}}
  \caption{{\bf Gibbs free energy of BN polymorphs at zero pressure
      expressed as a function of temperature.}~A phase transition
    between the c--BN and h--BN$(AA')$ polymorphs is predicted to occur
    at $T_{\rm c \to h} = 335 \pm 30$~K. Temperature-induced volume
    expansion effects are appropriately taken into account (Methods and
    Supplementary Figure~2). The mass density of the two polymorphs at
    the transition temperature are indicated along with the
    corresponding experimental room-temperature values (within
    parentheses, taken from works \cite{knittle89,solozhenko95}).
    The shaded area indicates the
    $\pm 0.26$~kJ/mol numerical error in
    the RPA and vibrational free energy calculations
    which leads to the $\pm 30$~K
    error in $T_{\rm c \to h}$ indicated by the horizontal bar.}
  \label{figT}
\end{figure}
Figure~\ref{figT} shows the Gibbs free energy $G$ of several BN
polymorphs estimated at zero pressure with the quasi-harmonic approach
(Methods). Static/vibrational contributions to $G$ are
calculated with the RPA/LDA method (Methods). Temperature-induced
volume expansion
effects are fully accounted for in our results in order to provide
precise phase transition data (Methods and Supplementary Figure~2). We
find that the h--BN$(AA')$ polymorph becomes entropically stabilized
over c--BN at $T_{\rm c \to h} = 335$~K, a temperature
relatively close to ambient conditions that falls significantly below
the corresponding experimental data
\cite{bundy62,solozhenko99,will00}. Taking into account a possible
numerical error of 0.2~kJ/mol in the RPA calculations, and an 
additional 0.06~kJ/mol in the vibrational free energies (Methods) leads 
to $T_{\rm c \to h}=335\pm 30$~K, still below the lowest
experimental result. We note, however, that the mass
densities that we estimate (using FI + LDA corrections) for
h--BN$(AA')$ and c--BN at $T_{\rm c \to h}$ agree almost perfectly
with the experimental measurements carried out at room temperature
(Fig.~\ref{figT}) \cite{knittle89,solozhenko95}.  The low-energy phonon
excitations in the h--BN$(AA')$ phase present much lower frequencies
than in c--BN (Supplementary Figure~3), hence the vibrational entropy
of the former polymorph becomes increasingly more favourable as the
temperature is raised.  Meanwhile, the Gibbs free energy of the
wurtzite polymorph, w--BN, falls out of the energy range considered in
Fig.~\ref{figT} due to its extreme vibrational stiffness, which
translates into destabilizing entropy contributions under increasing
temperature. By contrast, the Gibbs free energy of the m--BN,
h--BN$(AB)$, and h--BN$(ABC)$ (also known as r--BN) polymorphs follow
closely that of h--BN$(AA')$, falling all within an energy difference
range of about $1$~kJ/mol at $300 \le T \le 400$~K.

Our results confirm that the c--BN polymorph is most stable at low
pressures; however, the transition temperature that we predict for the
c--BN$\leftrightarrow$h--BN transformation lies relatively close to
ambient conditions. Consequently, the phase diagram of BN, if not
analogous, certainly is more similar than previously thought to that
of carbon. Actually, our $T_{\rm c \to h}$ estimation appears to be
consistent with the general belief based on empirical observations
that h--BN is most stable at ambient conditions. The likely reason for
the substantial difference between theory and measurements, $420 <
T^{\rm exp}_{\rm c \to h} < 1500$~K, may be the high-temperature 
high-pressure conditions and kinetic factors involved in the experimental
synthesis and analysis of BN samples \cite{will00}. We hope that our 
theoretical results will motivate new experimental activity in bulk BN.  
On the theory side, we have shown: i) that LDA is a good method for 
calculating elastic properties of materials, and hence is probably good 
for estimating vibrational free energies, but that ii) for accurate 
prediction of energy ordering among van der Waals polymorphs one must 
employ methods which include many-body dispersion interactions, ideally 
at a high-level using RPA, but certainly approximately when RPA is infeasible.


\section*{Methods}
{\bf Density functional theory and phonon
  calculations.}~First-principles calculations based on density
functional theory (DFT) are performed to analyze the energy,
structural, and vibrational properties of BN polymorphs. We perform
these calculations with the VASP code \cite{vasp}, using projector
augmented-wave method potentials \cite{bloch94}. The electronic states
$1s$-$2s$ of B and $2s$-$2p$ of N atoms are considered as
valence. Wave functions are represented in a plane-wave basis
truncated at $650$~eV. By using these parameters and dense
${\bf k}$-point grids for integration within the first Brillouin zone
(IBZ), energies are converged to within $1$~meV per formula unit
(0.1~kJ/mol, Supplementary Figure~4). In the geometry relaxations, a tolerance of
$0.01$~eV$\cdot$\AA$^{-1}$ is imposed in the atomic forces.

\emph{Ab initio} phonon frequencies are
calculated with the direct method in order to assess the vibrational
stability of the analyzed BN polymorphs and estimate their Gibbs free
energies as a function of temperature and pressure within the
quasi-harmonic approach \cite{cazorla17}. In the direct method the
force-constant matrix is calculated in real-space by considering the
proportionality between atomic displacements and forces
\cite{alfe09}. The quantities with respect to which our phonon
calculations are converged include the size of the supercell, the size
of the atomic displacements, and the numerical accuracy in the
sampling of the IBZ. We find the following settings to provide
quasi-harmonic free energies converged to within 0.1~kJ/mol:
$4 \times 4 \times 4$ supercells (where the figures indicate the
number of replicas of the unit cell along the corresponding lattice
vectors, Supplementary Figure~4), atomic displacements of $0.02$~\AA, and ${\bf q}$-point
grids of $14 \times 14 \times 14$. The value of the phonon frequencies
are obtained with the PHON code developed by Alf\`e \cite{alfe09}. In
using this code we exploit the translational invariance of the system,
to impose the three acoustic branches to be exactly zero at the center
of the Brillouin zone, and apply central differences in the atomic
forces.\\

{\bf Random phase approximation calculations.}~To overcome the
accuracy barrier of semi-empirical theories we carried out RPA
calculations, which give comparable results to high-level 
coupled-cluster and related wave function theories but are valid for
bulk systems with small or zero gaps \cite{rpa0,rpa1,rpa2,rpa3,rpa4}. Due
to its high numerical cost, our RPA calculations were performed using
structures optimised at the FI level \cite{fi}. Two sets of
calculations were carried out. Firstly, exact exchange (EXX) and RPA
correlation energy calculations,
\begin{align}
  E_{\rm RPA}=&E_{\rm EXX}^{\rm dense}@{\rm PBE}
  + E_{\rm c,RPA}^{\rm coarse}@{\rm PBE},
\end{align}
were carried out for all low-energy structures. These used an energy
cutoff of 480~eV, a dense $12\times 12\times 12$ (or equivalent)
${\bf k}$-point grid for EXX, and a coarser $7 \times 7 \times 7$ one
for RPA, evaluated on self-consistent PBE orbitals as per
standard practice \cite{rpa0}.
To further refine the energies of the c--BN, h--BN($AA'$) and m--BN
structures we carried out additional calculations using a more
accurate cutoff of 550~eV and the dense grid for both EXX and RPA;
this procedure yields energy difference results within 2~meV per formula 
unit (0.2~kJ/mol) of the initial calculations, which we use as our
numerical error bar. Errors 
in energy differences between various h--BN phases, and the m--BN phase
are expected to be much smaller ($< 0.2$~meV), due to the similarity of
the systems and consequent additional error cancellations.\\

{\bf Estimation of thermodynamic quantities.}~We use the
quasi-harmonic approach (QHA) \cite{cazorla17} to calculate the Gibbs
free energy $G$ of BN polymorphs as a function of temperature and
pressure. Anharmonic effects beyond the QHA have been shown to be
negligible for bulk BN at temperatures close to ambient conditions
(i.e., below 0.1~kJ/mol) \cite{kresse99}, hence we disregard them 
here. (We should note that in the unlikely case that anharmonicity 
played a role at $T \sim 300$~K, it would probably tend to further 
stabilise the hexagonal polymorph over the cubic \cite{kresse99}, 
thus additionally reducing $T_{\rm c \to h}$.)
In the QHA approximation the vibrational free energy of a crystal 
$F_{\rm vib}$ with volume $V$ and at temperature $T$ is:
\begin{equation}
  F_{\rm vib}(V,T) = \frac{1}{N_{q}}~k_{B} T \sum_{\boldsymbol{q}s}
  \ln\left[ 2\sinh \left( 
    \frac{\hbar\omega_{\boldsymbol{q}s}}{2k_{\rm B}T} \right) \right]~,
\label{eq:fharm}
\end{equation}
where $N_{q}$ is the total number of wave vectors used for integration
within the first Brillouin zone, the summation runs over all wave
vectors $\boldsymbol{q}$ and phonon branches $s$, and
$\omega_{\boldsymbol{q}s}$ are the vibrational frequencies of the
crystal, which depend on volume. In the zero-temperature limit
$F_{\rm vib}$ becomes:
\begin{equation}
  E_{\rm ZPE} = \frac{1}{N_{\rm q}} \sum_{\boldsymbol{q}s}
  \frac{1}{2}\hbar\omega_{\boldsymbol{q}s}~,
  \label{eq:zpe}
\end{equation}
which usually is referred to as the ``zero-point energy'' (ZPE). The
Gibbs free energy of a crystal then reads:
\begin{equation}
  G (V,T) = E_{\rm el} (V) + F_{\rm vib} (V,T) + PV~,
  \label{eq:gibbs}
\end{equation}
where $E_{\rm el}$ is the energy of the system when all atoms rest
immobile in their equilibrium positions, and the hydrostatic pressure
$P$ is estimated via the volume derivative:
\begin{equation}
  P (V,T) = -\left( \frac{\partial E_{\rm el}}{\partial V}
  + \frac{\partial F_{\rm vib}}{\partial V} \right)~.
  \label{eq:press}
\end{equation}

Finally, by using the thermodynamic constraint $P(V_{0},T) = 0$ and
performing $E_{\rm el}$ and $F_{\rm vib}$ calculations over a dense
grid of volume points, it is possible to account precisely for
$T$--induced volume expansion effects at zero pressure (Supplementary
Figure~2). The zero-temperature energies reported in this study
account for possible quantum nuclear effects by means of the
expression:
\begin{equation}
  E^{m} (V^{m}_{0}) = E^{m}_{\rm el} (V^{m}_{0}) 
  + E^{\rm LDA}_{\rm ZPE} (V^{\rm LDA}_{0})~,
\label{eq:zerotener}
\end{equation}
where ``$m$'' denotes the method of calculation, $V^{m}_{0}$ the
resulting equilibrium volume, and $E^{m}_{\rm ZPE}$ the zero-point
energy as given by Eq.(\ref{eq:zpe}). We have checked that the value
of zero-point energy differences between BN polymorphs are
practically independent of the employed method (Supplementary
Figure~5), hence the reason for our fixed choice of $E_{\rm ZPE}$ in
Eq.(\ref{eq:zerotener}).
In the RPA case, given the huge computational expense associated with
this method, the Gibbs free energies have been estimated by using both
$F_{\rm vib}$ and hydrostatic pressure values obtained with the LDA
method.

This particular choice is justified by the fact that LDA often
performs similarly to RPA for stress tensors \cite{leconte17},
as we have explicitly corroborated in this study (Table~\ref{tab1}).
  Based on numerical errors of 6\% for the elastic constants of LDA,
  versus RPA, we assign a corresponding numerical error to the
  vibrational free energies of $\pm 0.06~$kJ/mol (calculated as
  6\% of the zero-point energy difference between the cubic and
  hexagonal BN polymorphs, Supplementary Methods).

  We must note, however, that this good agreement is
  a result of LDA's ability to
  predict the energies of small crystal perturbations, as required
  for phonon calculations and elastic coefficients. It does not
  transfer to the energies of \emph{structurally distinct} systems
  required for accurate polymorph prediction, which explains
  LDA's failures in that regard (Supplementary Table~2).

\section*{Supplementary Materials}
Supplementary material for this article is available at \textit{xxx}.\\
table S1. Numerical tests performed for the RPA calculations.\\
table S2. Zero-temperature electronic energies of several BN polymorphs calculated with different first-principles methods.\\
table S3. Structural properties of the monoclinic phase m--BN reported in this study.\\
table S4. Elastic constants associated with compressive deformations calculated with the LDA method for different BN polymorphs.\\
fig. S1. Phonon spectrum of the new monoclinic phase m--BN calculated with the LDA method.\\
fig. S2. Gibbs free energy differences among several BN polymorphs calculated at zero-pressure and expressed as a function of temperature.\\
fig. S3. Phonon spectrum of the c--BN and h--BN polymorphs calculated with the LDA method.\\
fig. S4. Convergence tests of the electronic and vibrational free energies calculated with DFT methods for the h--BN polymorph.\\
fig. S5. Zero-point energy corrections for several BN polymorphs calculated with different DFT methods.\\

\section*{Acknowledgments}
{\bf Funding:} This research was supported under the Australian Research Council's
Future Fellowship funding scheme (No. FT140100135). Computational
resources and technical assistance were provided by the Australian
Government and the Government of Western Australia through Magnus
under the National Computational Merit Allocation Scheme and The
Pawsey Supercomputing Centre, and by Gowonda high-performance
computing facilities. {\bf Author contributions:} C.C. and T.G. designed the research. 
T.G. carried out DFT and RPA energy calculations. C.C. carried out the structural 
and thermodynamic analysis. C.C. and T.G. wrote the manuscript. Both authors 
contributed equally to the present work. {\bf Competing interests:} The authors declare 
no competing interests. {\bf Data availability:} All data needed to evaluate the conclusions 
in the study are present in the paper and/or the Supplementary Materials. The data that 
support the findings of this study are available from the corresponding authors (C.C. and T.G.) 
upon reasonable request.


\begin{thebibliography}{30}
\bibitem{dean10}  Dean, C. R., Young, A. F., Meric, I., Lee, C., Wang, L., Sorgenfrei, S., Watanabe, K., 
	          Taniguchi, T., Kim, P., Shepard, K. L. $\&$ Hone, J.
		  Boron nitride substrates for high-quality graphene electronics.
		  \textit{Nat. Nanotechnol.} \textbf{5}, 722 (2010).

\bibitem{tang02} Tang, C. C., Bando, Y., Sato, T. $\&$ Kurashima, K.
	         Uniform boron nitride coatings on silicon carbide nanowires.
		 \textit{Adv. Mater.} \textbf{14}, 1046 (2002).

\bibitem{pease50} Pease, R. S.
                  Crystal structure of boron nitride.
                  \textit{Nature} \textbf{165}, 722 (1950).

\bibitem{corrigan75} Corrigan, F. R. $\&$ Bundy, F. P.
                     Direct transitions among the allotropic forms of boron nitride at high pressures and temperatures.
		     \textit{J. Chem. Phys.} \textbf{63}, 3812 (1975).

\bibitem{narayan16} Narayan, J., Bhaumik, A. $\&$ Xu, W.
                    Direct conversion of h--BN into c--BN and formation of epitaxial c--BN/diamond heterostructures.
                    \textit{J. Chem. Phys.} \textbf{119}, 185302 (2016). 

\bibitem{bundy62} Bundy, F. P. $\&$ Wentorf, R. H.
                  Direct transformation of hexagonal boron nitride to denser forms.
                  \textit{J. Chem. Phys.} \textbf{38}, 1144 (1963).

\bibitem{solozhenko99} Solozhenko, V. L., Turkevich, V. Z. $\&$ Holzapfel, W. B.
                       Refined phase diagram of boron nitride.
                       \textit{J. Phys. Chem. B} \textbf{103}, 2903 (1999). 

\bibitem{will00} Will, G., Nover, G. $\&$ von der G\"{o}nna, J. 
                 New experimental results on the phase diagram of boron nitride.
                 \textit{J. Solid State Chem.} \textbf{154}, 280 (2000).

\bibitem{lda} Ceperley, D. M. $\&$ Alder, B. J.
              Ground state of the electron gas by a stochastic method.
              \textit{Phys. Rev. Lett.} \textbf{45}, 566 (1980).

\bibitem{pbe} Perdew, J. P., Burke, K. $\&$ Ernzerhof, M.
              Generalized gradient approximation made simple.
              \textit{Phys. Rev. Lett.} \textbf{77}, 3865 (1996).

\bibitem{kresse99} Kern, G., Kresse, G. $\&$ Hafner, J.
                   Ab initio calculation of the lattice dynamics and phase diagram of boron nitride.
                   \textit{Phys. Rev. B} \textbf{59}, 8551 (1999).

\bibitem{ahmed07} Ahmed, R., Fazal-e-Aleem, J., Hashemifar, S. J. $\&$ Akbarzadeh, H.
                  First principles study of structural and electronic properties of different phases of boron nitride.
                  \textit{Physica B} \textbf{400}, 297 (2007).

\bibitem{d31} Grimme, S., Antony, J., Ehrlich, S. $\&$ Krieg, H.
              A consistent and accurate ab initio parametrization of density functional dispersion correction (DFT-D) 
	      for the 94 elements H-Pu.  
              \textit{ J. Chem. Phys.} \textbf{132}, 154104 (2010).  

\bibitem{d32} Goerigk, L. $\&$ Grimme, S. 
              A thorough benchmark of density functional methods for general main group thermochemistry, kinetics, 
	      and noncovalent interactions.
              \textit{ Phys. Chem. Chem. Phys.} \textbf{13}, 6670 (2011).

\bibitem{d3bj} Grimme, S., Ehrlich, S. $\&$ Goerigk, L.
               Effect of the damping function in dispersion corrected density functional theory.
               \textit{J. Comput. Chem.} \textbf{32}, 1456 (2011).

\bibitem{mbd} Tkatchenko, A., DiStasio, R. A., Car, R. $\&$ Scheffler, M.
              Accurate and efficient method for many--body van der Waals interactions.
              \textit{Phys. Rev. Lett.} \textbf{108}, 236402 (2012).

\bibitem{fi} Gould, T., Leb\`{e}gue, S., \'{A}ngy\'{a}n, J. G. $\&$ Bu\v{c}ko, T.
             A fractionally ionic approach to polarizability and van der Waals many-body dispersion calculations. 
             \textit{J. Chem. Theor. Comput.} \textbf{12}, 5920 (2016).

\bibitem{tawfik2018} Tawfik, S. A., Gould, T, Stampfl C. $\&$ Ford M. J.
               Evaluation of van der Waals density functionals for layered materials.
               \textit{Phys. Rev. Materials} \textbf{2}, 034005 (2018).

\bibitem{ambrosetti2016} Ambrosetti, A., Ferri. N., DiStasio Jr., R. A. $\&$ Tkatchenko, A.
               Wavelike charge density fluctuations and van der Waals interactions at the nanoscale.
               \textit{Science} \textbf{351}, 1171 (2016).

\bibitem{hermann2017} Hermann, J., DiStasio Jr., R. A. $\&$ Tkatchenko, A.
               First-principles models for van der Waals interactions in molecules and materials: concepts, theory, and applications.
               \textit{Chem. Rev.} \textbf{117}, 4714 (2017).

\bibitem{dobson2014} Dobson, J. F.
               Beyond pairwise additivity in London dispersion interactions.
               \textit{Int. J. Quant. Chem.} \textbf{114}, 1157 (2014).

\bibitem{rpa0} Harl, J. $\&$ Kresse, G.
               Accurate bulk properties from approximate many-body techniques.
               \textit{Phys. Rev. Lett.} \textbf{103}, 056401 (2009).

\bibitem{rpa1} Dobson, J. F. $\&$ Gould, T.
               Calculation of dispersion energies.
               \textit{J. Phys.: Condens. Matt.} \textbf{24}, 073201 (2012).  

\bibitem{rpa2} Eshuis, H., Bates, J. E. $\&$ Furche, F.
               Electron correlation methods based on the random phase approximation.
               \textit{Theor. Chem. Acc.} \textbf{131}, 1 (2012).

\bibitem{rpa3} Ren, X., Rinke, P., Joas, C. $\&$ Scheffler, M. 
	       Random--phase approximation and its applications in computational chemistry and materials science.
               \textit{J. Mater. Sci.} \textbf{47}, 7447 (2012).

\bibitem{rpa4} Chen, G. P., Voora, V. K., Agee, M. M., Balasubramani, S. G. $\&$ Furche, F.
	       Random--phase approximation methods.
	       \textit{Annu. Rev. Phys. Chem.} \textbf{68}, 421 (2017).

\bibitem{cazorla17} Cazorla, C. $\&$ Boronat, J.
                    Simulation and understanding of atomic and molecular quantum crystals.
                    \textit{Rev. Mod. Phys.} \textbf{89}, 035003 (2017).

\bibitem{zhang13} Zhang, X., Wang, Y., Lv, J., Zhu, C., Li, Q., Zhang, M., Li, Q. $\&$ Ma, Y.
                  First-principles structural design of superhard materials. 
                  \textit{J. Chem. Phys.} \textbf{138}, 114101 (2013).

\bibitem{horiuchi95} Horiuchi, S., He, L.-L., Onoda, M. $\&$ Akaishi, M.
                     Monoclinic phase of boron nitride appearing during the hexagonal cubic phase transition at high 
                     pressure and high temperature.
                     \textit{Appl. Phys. Lett.} \textbf{68}, 182 (1995). 

\bibitem{ivanovskii12} Ivanovskii, A. L. 
                       Mechanical and electronic properties of diborides of transition $3d$--$5d$ metals from 
                       first principles: Toward search of novel ultra-incompressible and superhard materials.
                       \textit{Prog. Mater. Sci.} \textbf{57}, 184 (2012).

\bibitem{knittle89} Knittle, E., Wentzcovitch, R. M., Jeanloz, R. $\&$ Cohen, M. L. 
                    Experimental and theoretical equation of state of cubic boron nitride.
                    \textit{Nature} \textbf{337}, 349 (1989).

\bibitem{solozhenko95} Solozhenko, V. L., Will, G. $\&$ Elf, F.
                       Isothermal compression of hexagonal graphite-like boron nitride up to 12 GPa.
                       \textit{Solid State Commun.} \textbf{96}, 1 (1995).

\bibitem{vasp} Kresse, G. $\&$ Furthm\"{u}ller, J.
               Efficient iterative schemes for ab initio total-energy calculations using a plane-wave basis set.
               \textit{Phys. Rev. B} \textbf{54}, 11169 (1996).

\bibitem{bloch94} Bl\"ochl, P. E.
                  Projector augmented-wave method.
                  \textit{Phys. Rev. B} \textbf{50}, 17953 (1994).

\bibitem{alfe09} Alf\`e, D.
                 PHON: A program to calculate phonons using the small displacement method.
                 \textit{Comp. Phys. Commun.} \textbf{180}, 2622 (2009).

\bibitem{leconte17} Leconte, N., Jung, J., Leb\`{e}gue, S. $\&$ Gould, T.
                    Moir\'{e}-pattern interlayer potentials in van der Waals materials in the random-phase approximation.
		    \textit{Phys. Rev. B} \textbf{96}, 195431 (2017).	

\bibitem{li15} Li, Y., Hao, J., Liu, H., Lu, S. $\&$ Tse, J. S.
               High-energy density and superhard nitrogen-rich B-N compounds
               \textit{Phys. Rev. Lett.} \textbf{115}, 105502 (2015).

\bibitem{isotropy} Stokes, H. T. $\&$ Hatch, D. M.
                   FINDSYM: program for identifying the space-group symmetry of a crystal.
                   \textit{J. Appl. Cryst.} \textbf{38}, 237 (2005).

\end{thebibliography}
\end{document}